\documentclass[aps,preprintnumbers,superscriptaddress,showpacs]{revtex4}
\usepackage{epsfig}
\usepackage{psfrag}
\usepackage{amsfonts}
\usepackage{graphicx}
\usepackage{dcolumn}
\usepackage{bm}
\usepackage{amsmath}
\usepackage{relsize}

\begin{document}

\title{Gluon condensate effects on heavy quarkonium spectral functions and thermal dissociation}

\author{Fei Wang}
\affiliation{School of Mathematics and Physics, China University
of Geosciences (Wuhan), Wuhan 430074, China}

\author{Zi-qiang Zhang}
\email{zhangzq@cug.edu.cn} \affiliation{School of Mathematics and
Physics, China University of Geosciences (Wuhan), Wuhan 430074,
China}

\begin{abstract}
We investigate how the gluon condensate modifies the thermal
spectral functions and melting patterns of heavy vector mesons,
specifically charmonium and bottomonium, using an improved
soft-wall AdS/QCD model. The framework is extended to finite
temperature via a dilaton black hole geometry that consistently
incorporates the backreaction from the gluon condensate. We
numerically track the evolution of spectral resonance peaks as
functions of both temperature and gluon condensate strength. Our
calculations reveal that increasing temperature systematically
broadens and suppresses spectral peaks, signaling in-medium
dissociation. In contrast, a stronger gluon condensate
considerably mitigates peak broadening and enhances the spectral
weight of both ground and excited states. This behavior indicates
that the gluon condensate hinders thermal dissociation, thereby
stabilizing heavy quarkonia within the quark-gluon plasma. These
findings are consistent with existing studies and provide new
holographic evidence for the stabilising role of the gluon
condensate from the perspective of thermal spectral functions.
\end{abstract}

\maketitle
\section{Introduction}

Heavy vector mesons serve as essential probes for characterizing
the quark-gluon plasma (QGP) generated in ultrarelativistic
heavy-ion collisions. Their experimentally observed suppression is
commonly attributed to in-medium dissociation inside the hot
medium~\cite{TM1986,HS2006}, a key signature used to identify
deconfined QCD matter. Therefore, understanding the thermodynamic
properties of heavy quarkonia and determining their dissociation
temperatures is crucial for clarifying the evolution of strongly
interacting matter under extreme conditions. Over the past almost
thirty years, numerous holographic studies have explored the
properties of heavy mesons using gauge/gravity
duality~\cite{MA2004,SH2004,YK2007_PRD,MF2009,JN2009,MF2010,TB2010,TG2013,KH2015,YL2017}.

Bottom-up AdS/QCD models have become a practical tool for
investigating heavy vector mesons in strongly coupled
plasmas~\cite{NRF2016,NRF2017,NRF2017_2,NRF2018,NRF2018_2,MAS2019,NRF2020,NRF2023,ZRZ2024,ZRZ2025,YQZ2022,LAH2014,XLW2024,WBC2024}.
In particular, an improved soft-wall holographic framework was
constructed to describe charmonium and bottomonium states at
finite temperature and density~\cite{NRF2016,NRF2017}. Built upon
the AdS/CFT correspondence~\cite{JMM1999,SSG1998,EW1998}, this
model introduces a background scalar field $\chi(z)$ with multiple
independent energy scales. With suitable parameters, it
simultaneously reproduces the vacuum mass spectrum and decay
constants of heavy vector mesons, and admits consistent finite
temperature and density generalizations. Unlike the original
hard-wall~\cite{JP2002,NRF2002,NRF2003,NRF2005} and soft-wall
models~\cite{AK2006}, the improved formulation naturally accounts
for the experimentally observed monotonic decrease of decay
constants with increasing radial excitation number. This advantage
makes it a consistent platform for studying heavy meson
dissociation in thermal environments.

Within this holographic setting, spectral functions provide a
direct window into quarkonium dissociation in the QGP, as they are
tightly connected to vector current two-point correlators. At zero
temperature, the correlator obeys the spectral decomposition
\begin{equation}
    \Pi(p^2) = \sum_{n=1}^{\infty} \frac{f_n^2}{(-p^2) - m_n^2 + i\epsilon}.
\end{equation}
Evaluating the left-hand side holographically allows extraction of
meson masses and decay constants~\cite{AK2006,HRG2007}. The
imaginary part reduces to a sum of delta functions proportional to
$f_n^2\,\delta(-p^2 - m_n^2)$, demonstrating that decay constants
determine the strength of spectral resonances at zero temperature.
At finite temperature, the sharp delta peaks broaden into
structures with finite width. The position and shape of these
resonances encode information about quasiparticle masses,
lifetimes, and dissociation. Consequently, thermal spectral
functions, defined as the imaginary part of retarded Green's
functions, occupy a central position in the analysis of heavy
vector mesons in hot media. The holographic construction enables a
smooth extension from zero to finite temperature, providing a
solid foundation for analyzing quarkonium melting via spectral
functions or quasinormal modes.

The gluon condensate represents a key nonperturbative feature of
the QCD vacuum and significantly influences the phase structure
and dynamics of quantum chromodynamics~\cite{g3}. Within
holography, gluon condensate effects can be incorporated
self-consistently through backreaction in a gravity-dilaton
system. We denote the dilaton field responsible for this
backreaction as $\phi(z)$, to avoid confusion with the vacuum
soft-wall scalar $\chi(z)$ introduced above. These two scalars
originate from distinct physical ingredients: $\chi(z)$ encodes
the hierarchy of masses and decay constants for vacuum heavy
quarkonia, while $\phi(z)$ arises from the vacuum expectation
value of gluon operators and modifies the bulk black hole geometry
at finite temperature. Such geometric distortions alter the
thermodynamic response of the medium. In recent years, many
observables sensitive to the QGP have been studied in backgrounds
containing a gluon condensate, including heavy quark
potentials~\cite{YK2009}, jet quenching parameters~\cite{ZQZ2019},
Schwinger effect~\cite{ZQZ2020}, imaginary interquark
potentials~\cite{ST2023}, and entropic forces~\cite{ZQZ2020_2}.
Collectively, these works show that gluon condensation strongly
influences characteristic QGP properties and offers insight into
the nonperturbative behavior of strongly coupled quark-gluon
matter.

In previous work~\cite{wang2025}, we employed configurational
entropy (CE) to examine QGP stability in geometries with a gluon
condensate. CE quantifies system stability from an information
theoretic perspective~\cite{MG1,MG2,MG3}. Our results indicated
that larger gluon condensates lower the CE, signaling reduced
microscopic disorder and enhanced QGP stability, with clear
implications for the QCD deconfinement transition. Nevertheless,
the CE formalism mainly captures bulk thermodynamic stability. It
cannot resolve fine spectral features such as resonance broadening
or track dynamical processes like quarkonium thermal melting.
These microscopic properties are essential for understanding the
internal structure and evolution of the QGP and remain
inaccessible to purely entropic analyses. Thermal spectral
functions, by contrast, directly reveal medium modifications of
heavy vector mesons at finite temperature and density, making them
an indispensable tool for probing QGP microphysics. Therefore, a
systematic study of heavy meson spectral functions in holographic
backgrounds with a gluon condensate naturally complements and
extends our earlier CE work, bridging the gap between macroscopic
stability arguments and microscopic quasiparticle dynamics.

The layout of this paper is as follows. Section \ref{sec:softwall}
reviews the improved soft-wall model for heavy vector mesons and
summarizes the relation between decay constants and spectral
peaks. Section \ref{sec:dilatonbh} extends the setup to a dilaton
black hole geometry carrying a gluon condensate, where the
geometric scalar field $\phi(z)$ is distinguished from the vacuum
soft-wall scalar $\chi(z)$. Section \ref{sec:membrane} outlines
the derivation and numerical evaluation of vector meson spectral
functions using the membrane paradigm. Section \ref{sec:numerics}
presents numerical results for charmonium and bottomonium thermal
spectra, varying temperature and gluon condensate strength.
Section \ref{sec:summary} contains our conclusions and outlook.

\section{Improved soft wall Model for Heavy Mesons}
\label{sec:softwall} We start by recalling the connection between
decay constants and spectral peaks within the holographic
framework~\cite{NRF2016,NRF2017}. Heavy vector mesons are
described by a five-dimensional vector field
\begin{equation}
    V_m = (V_\mu, V_z), \qquad \mu = 0,1,2,3,
\end{equation}
which is dual under gauge/gravity correspondence to the conserved
vector current of the boundary gauge theory:
\begin{equation}
    J^\mu = \bar{\psi}\gamma^\mu\psi.
\end{equation}

The bulk dynamics follow from the action
\begin{equation}
    I = \int d^4x\,dz\,\sqrt{-g}\,e^{-\chi(z)}
    \left(
    -\frac{1}{4g_5^2} F_{mn}F^{mn}
    \right),
\end{equation}
with field strength
\begin{equation}
    F_{mn} = \partial_m V_n - \partial_n V_m.
\end{equation}
Here $g_5$ is the five-dimensional gauge coupling, and $\chi(z)$
is the vacuum soft-wall scalar encoding nonperturbative QCD
effects relevant for zero-temperature hadron spectra. At zero
temperature the background geometry is five-dimensional AdS space,
\begin{equation}
    ds^2 = \frac{R^2}{z^2}
    \left( -dt^2 + d\vec{x}\cdot d\vec{x} + dz^2 \right),
\end{equation}
where $z$ denotes the radial coordinate and $R$ the AdS radius.
Working in radial gauge $V_z = 0$, the four-dimensional components
$V_\mu$ act as sources for boundary current correlators.

To simultaneously reproduce the mass spectrum and decay constants
of heavy vector mesons, we adopt an improved soft-wall scalar
profile built from three independent energy scales:
\begin{equation}
    \chi(z) = k^2 z^2 + M z
    + \tanh\!\left( \frac{1}{Mz} - \frac{k}{\sqrt{\Gamma}} \right).
\end{equation}
Its ultraviolet and infrared asymptotic behavior clarifies how it
generates an effective geometric cutoff. In the UV limit $z\to 0$,
the argument inside the hyperbolic tangent becomes large positive,
so $\tanh(x)\to 1$, yielding $\chi(z) \approx 1 + M z + k^2 z^2$.
This mild linear-quadratic growth at small $z$ modifies the UV
scaling of the bulk gauge kinetic term and reproduces the
experimentally observed ordering of decay constants. In the IR
region $z\to\infty$, the argument approaches the constant
$-k/\sqrt{\Gamma}$, so the tanh term saturates to a fixed negative
value, while the quadratic term $k^2 z^2$ dominates and produces a
smooth infrared confinement scale without the sharp discontinuity
present in hard-wall models. The parameter $k$ relates to the
heavy quark mass, $\Gamma$ characterizes the string tension of the
$Q\bar{Q}$ interaction, and $M$ corresponds to the mass scale
relevant for non-hadronic decay channels. For charmonium and
bottomonium we use the fitted parameters from
Refs.~\cite{NRF2016,NRF2017,NRF2017_2,NRF2018,NRF2018_2,MAS2019,NRF2020,NRF2023}:
\begin{align}
    k_c &= 1.2~\mathrm{GeV}, \quad \sqrt{\Gamma_c} = 0.55~\mathrm{GeV}, \quad M_c = 2.2~\mathrm{GeV} \label{cc}, \\
    k_b &= 2.45~\mathrm{GeV}, \quad \sqrt{\Gamma_b} = 1.55~\mathrm{GeV}, \quad M_b = 6.2~\mathrm{GeV} \label{bb}.
\end{align}

This scalar profile implements a smooth infrared cutoff and
introduces necessary UV corrections, allowing the model to capture
the monotonic decrease of heavy vector meson decay constants with
radial excitation number.

Setting $V_z = 0$, the momentum-space equations of motion for
transverse vector field components (collectively denoted $V$) read
\begin{equation}
    \partial_z \!\left[ e^{-B(z)} \partial_z V \right]
    - p^2 e^{-B(z)} V = 0 ,
\end{equation}
where
\begin{equation}
    B(z) = \log\!\left( \frac{z}{R} \right) + \chi(z) .
\end{equation}

Normalizable solutions correspond to discrete meson states. The
eigenvalues
\begin{equation}
    p^2 = - m_n^2
\end{equation}
determine the meson mass spectrum, while decay constants are
extracted from the near-boundary behavior of the mode functions
$\Psi_n(z)$~\cite{NRF2017_2}:
\begin{equation}
    f_n = \frac{1}{g_5 m_n} \lim_{z \to 0} \left( \frac{R}{z} e^{-\chi(z)} \partial_z \Psi_n(z) \right).
\end{equation}
This relation follows directly from the standard holographic
dictionary for vector-current two-point functions, where the
boundary residue of bulk modes maps onto QCD decay constants;
detailed matching for soft-wall AdS/QCD can be found in
Refs.~\cite{AK2006,HRG2007}. As already noted, the
zero-temperature spectral decomposition shows that decay constants
directly control the strength of quasiparticle peaks in spectral
functions. The model therefore offers a consistent holographic
framework for investigating heavy vector meson dissociation in
finite temperature, finite density, and external field
backgrounds.

\section{AdS Black Hole with Gluon Condensate}
\label{sec:dilatonbh} We now extend the holographic setup to
include gluon condensate effects. Within gauge/gravity duality,
these effects are realized via a five-dimensional gravity-dilaton
system dual to pure gluodynamics. We use $\phi(z)$ for the bulk
scalar driving gluon condensate backreaction, keeping it separate
from the vacuum soft-wall scalar $\chi(z)$ defined in
Sec.~\ref{sec:softwall}. The action is taken
as~\cite{SN1999,ZRZ2022,YK2007}:
\begin{equation}
    I=-\frac{1}{2\kappa^2}\int d^5x\sqrt{g}
    \left(R+\frac{12}{L^2}-\frac{1}{2}\partial_\mu\phi\,\partial^\mu\phi\right),
\end{equation}
where $\kappa^2$ is the five-dimensional gravitational coupling,
$R$ the Ricci scalar, $L$ the AdS radius, and $\phi$ the dilaton
dual to the gluon operator in the boundary theory. Dilaton
backreaction breaks the conformal invariance of the AdS$_5$
geometry and encodes gluon condensate effects holographically.

At finite temperature the system supports a dilaton black hole
solution~\cite{YK2007,AK1999,CC2007,ZRZ2022} with metric
\begin{equation}
    ds^2=\frac{1}{z^2}\left[X(z)\,d\vec{x}^2-Y(z)\,dt^2+dz^2\right],
\end{equation}
where
\begin{align}
    X(z)&=\left(1+f z^4\right)^{\frac{f+a}{2f}}
    \left(1-f z^4\right)^{\frac{f-a}{2f}}, \nonumber\\
    Y(z)&=\left(1+f z^4\right)^{\frac{f-3a}{2f}}
    \left(1-f z^4\right)^{\frac{f+3a}{2f}},
\end{align}
and the parameter $f$ satisfies
\begin{equation}
    f^2=a^2+c^2.
\end{equation}

The dilaton profile reads
\begin{equation}
    \phi(z)=\frac{c}{f}\sqrt{\frac{3}{2}}
    \log\!\left(\frac{1+f z^4}{1-f z^4}\right)+\phi_0.
\end{equation}
The constant $c$ controls the magnitude of the gluon condensate
and is proportional to the expectation value of the gluon field
strength squared,
\begin{equation}
    \langle G_{\mu\nu}^2 \rangle =
    \frac{8\sqrt{3(N_c^2-1)}}{\pi}\,c .
\end{equation}
It therefore serves as a measure of the nonperturbative gluonic
background~\cite{ZRZ2024}. In the limit $c\to0$, the geometry
reduces to the standard AdS black hole, while nonzero $c$ modifies
both the metric and dilaton profile.

The event horizon sits at $fz^4=1$. The Hawking temperature
follows from regularity of the Euclidean metric at the horizon.
Although a simple conical singularity argument for temperature
does not apply when $c\neq0$, the parameter $a$ can still be
related to the boundary theory temperature via $a={(\pi
T)^4}/{4}$~\cite{ZQZ2019}. This dilaton black hole geometry with
gluon condensate provides a natural holographic background to
explore thermal properties and stability of the QGP.

\section{Spectral Function via Membrane Paradigm}
\label{sec:membrane} We compute heavy vector meson spectral
functions in the gluon condensate background using the membrane
paradigm~\cite{NI2009}. Vector mesons are modeled by the
five-dimensional vector field $V_m=(V_\mu,V_z)$, holographically
dual to the conserved boundary current $J^\mu =
\bar{\psi}\gamma^\mu\psi$. We consider a general black brane
geometry of the form
\begin{equation}
    ds^2=-g_{tt}dt^2+g_{zz}dz^2+g_{x_1x_1}dx_1^2+g_{x_2x_2}dx_2^2+g_{x_3x_3}dx_3^2, \label{ds_2}
\end{equation}
with the AdS boundary at $z=z_0$ and the black hole horizon $z_h$
defined by $g_{tt}(z_h)=0$. The bulk gauge field obeys the action
\begin{equation}
    S=-\int d^5x\,\sqrt{-g}\,\frac{1}{4g_5^2\,h(z)}F^{mn}F_{mn},
\end{equation}
where $h(z)$ is a $z$-dependent effective coupling set to
$e^{\chi(z)}$ following earlier improved soft-wall
constructions~\cite{NRF2018}. Varying the action yields the
equations of motion
\begin{equation}
    \partial^m\!\left(\frac{\sqrt{-g}}{h(z)}F_{mn}\right)=0.
\end{equation}

On constant-$z$ hypersurfaces we define the gauge field conjugate
momentum
\begin{equation}
    j^\mu=-\frac{1}{h(z)}\sqrt{-g}\,F^{z\mu}. \label{jmu}
\end{equation}

We assume translational symmetry in the transverse directions so
$g_{x_1 x_1}=g_{x_2 x_2}$ and work with field configurations
independent of $x_1,x_2$. The equations of motion split into
longitudinal and transverse channels, leading to
\begin{align}
    -\partial_z j^t-\frac{\sqrt{-g}}{h}g^{tt}g^{x_3x_3}\partial_{x_3}F_{x_3t}&=0,\label{jt}\\
    -\partial_z j^{x_3}+\frac{\sqrt{-g}}{h}g^{tt}g^{x_3x_3}\partial_t F_{x_3t}&=0,\\
    \partial_{x_3}j^{x_3}+\partial_t j^t&=0 .  \label{jx3_jt}
\end{align}

Making use of the Bianchi identity we obtain
\begin{equation}
    \partial_z F_{x_3t}
    -\frac{h(z)}{\sqrt{-g}}g_{zz}g_{x_3x_3}\partial_t j^z
    -\frac{h(z)}{\sqrt{-g}}g_{tt}g_{x_3x_3}\partial_{x_3}j^t=0  \label{Bianchi}.
\end{equation}

Introduce a longitudinal $z$-dependent conductivity
\begin{equation}
    \bar{\sigma}_L(\omega,\vec{p},z)
    =\frac{j^{x_3}(\omega,\vec{p},z)}{F_{x_3t}(\omega,\vec{p},z)}.
\end{equation}
Differentiating with respect to $z$, substituting Eqs.~(\ref{jt}),
(\ref{jx3_jt}), (\ref{Bianchi}) and adopting the plane-wave ansatz
$p=(\omega,0,0,p_3)$, we arrive at the flow equation
\begin{equation}
    \partial_z \bar{\sigma}_L = - i \omega \sqrt{\frac{g_{zz}}{g_{tt}}} \left[ \Sigma(z) - \frac{\bar{\sigma}_L^2}{\Sigma(z)}
    \left( 1 - \frac{p_3^2}{\omega^2} \frac{g^{x_3 x_3}}{g^{tt}} \right) \right] \label{delta_L}.
\end{equation}
Here
\begin{equation}
    \Sigma(z) = \frac{1}{h(z)} \sqrt{\frac{-g}{g_{zz} g_{tt}}} \, g^{x_3 x_3}.
\end{equation}

An analogous treatment for the transverse sector~\cite{NI2009}
gives
\begin{equation}
    \partial_z \bar{\sigma}_T = i \omega \sqrt{\frac{g_{zz}}{g_{tt}}} \left[ \frac{\bar{\sigma}_T^2}{\Sigma(z)} - \Sigma(z)
    \left( 1 - \frac{p_3^2}{\omega^2} \frac{g^{x_3 x_3}}{g^{tt}} \right) \right].
\end{equation}

In the zero-momentum limit $p_3 \to 0$, both flow equations
simplify to identical form
\begin{equation}
    \partial_z \bar{\sigma} = i \omega \sqrt{\frac{g_{zz}}{g_{tt}}} \left[ \frac{\bar{\sigma}^2}{\Sigma(z)} - \Sigma(z) \right],
\end{equation}
with $\bar{\sigma}=\bar{\sigma}_L=\bar{\sigma}_T$. From the Kubo
relation $\sigma(\omega)= i G_R(\omega)/\omega$, the AC
conductivity reads
\begin{equation}
    \sigma(\omega) = - \frac{G_R(\omega)}{i \omega} = \bar{\sigma}(\omega,z_0).
\end{equation}

Since $\mathrm{Re}\,\sigma(\omega)=\rho(\omega)/\omega$, where
$\rho(\omega)\equiv \mathrm{Im}\,G_R(\omega)$ denotes the spectral
function, we apply the membrane paradigm to our geometry. With
metric (\ref{ds_2}) and $h(z)=e^{\chi(z)}$~\cite{NRF2018}, the
flow equation becomes
\begin{equation}
    \partial_z \bar{\sigma}(\omega,z) = \frac{i \omega}{Y(z)\,\bar{\Sigma}(z)} \left[ \bar{\sigma}(\omega,z)^2 - \bar{\Sigma}(z)^2 \right],
\end{equation}
where $\bar{\Sigma}(z) = \frac{X(z)^{1/2}}{z h(z)}$.

Horizon regularity imposes the boundary condition
$\bar{\sigma}(\omega,z_h)=\bar{\Sigma}(z_h)$. The spectral
function at the boundary is finally
\begin{equation}
    \rho(\omega) = \mathrm{Im}\,G_R(\omega) = \omega\,\mathrm{Re}\,\bar{\sigma}(\omega,z_0) \label{rho_omega}.
\end{equation}

\section{Numerical Results}
\label{sec:numerics} We solve the flow equations subject to the
horizon boundary condition and extract spectral functions. No
closed-form analytical solution exists, so we resort to numerical
integration. The parameter set follows the vacuum fits summarized
in Sec.~\ref{sec:softwall}: Eq.~(\ref{cc}) for charmonium and
Eq.~(\ref{bb}) for bottomonium.

\begin{figure}[h]
    \centering
    \includegraphics[width=8cm]{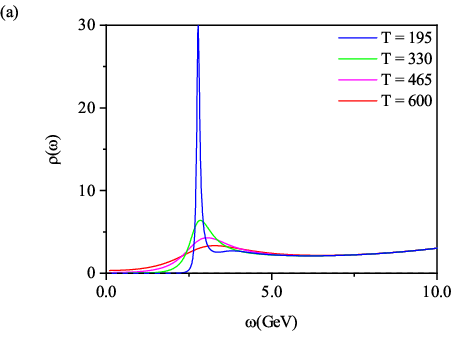}
    \includegraphics[width=8cm]{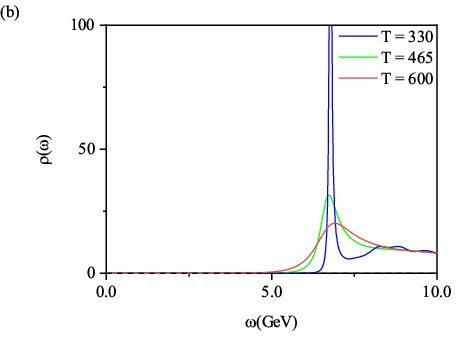}
    \caption{Spectral functions of charmonium (left panel) and bottomonium (right panel) at selected temperatures with vanishing gluon condensate $c=0$. Blue: $T=195\,\mathrm{MeV}$; green: $T=330\,\mathrm{MeV}$; pink: $T=465\,\mathrm{MeV}$; red: $T=600\,\mathrm{MeV}$.}
    \label{sp_t}
\end{figure}

\begin{figure}[h]
    \centering
    \includegraphics[width=8cm]{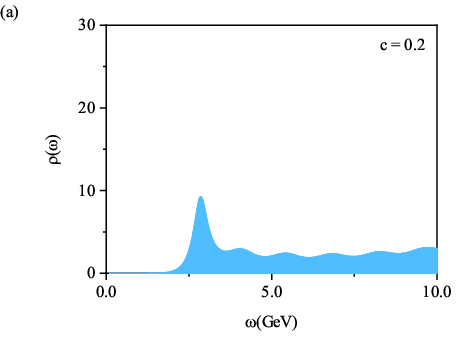}
    \includegraphics[width=8cm]{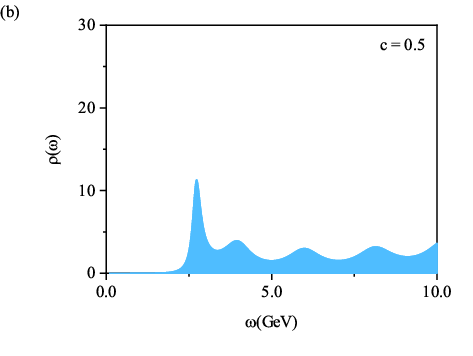}
    \includegraphics[width=8cm]{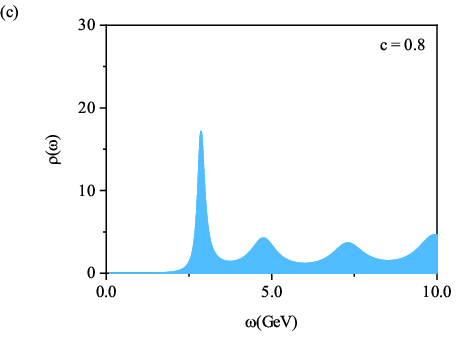}
    \includegraphics[width=8cm]{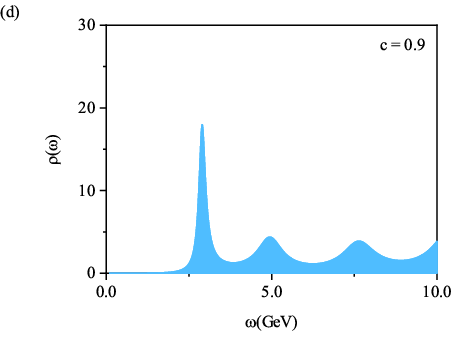}
    \caption{Charmonium spectral functions at fixed temperature $T=280$ MeV for different gluon condensate strengths: (a) $c=0.2$, (b) $c=0.5$, (c) $c=0.8$, (d) $c=0.9$.}
    \label{cc_c}
\end{figure}

\begin{figure}[h]
    \centering
    \includegraphics[width=8cm]{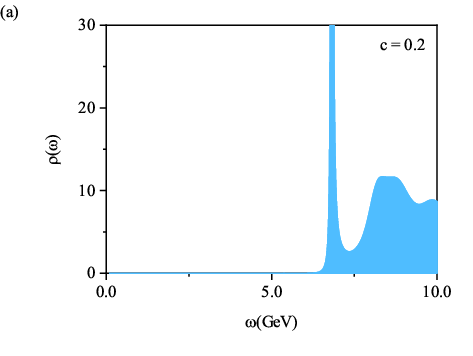}
    \includegraphics[width=8cm]{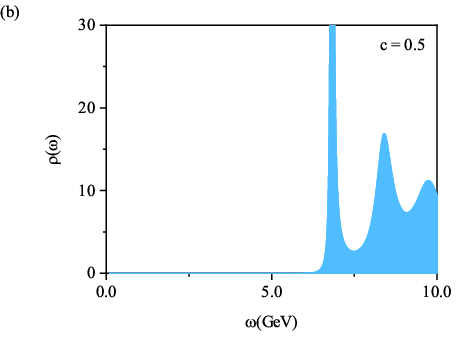}
    \includegraphics[width=8cm]{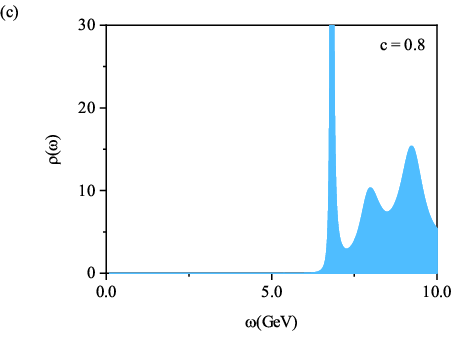}
    \includegraphics[width=8cm]{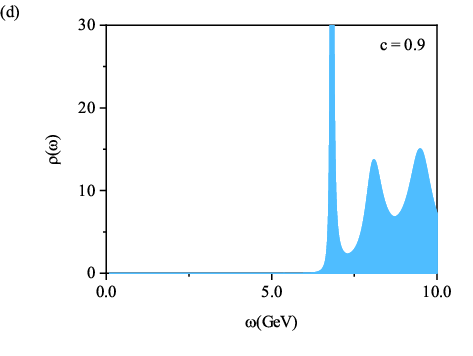}
    \caption{Bottomonium spectral functions at fixed temperature $T=300$ MeV for different gluon condensate strengths: (a) $c=0.2$, (b) $c=0.5$, (c) $c=0.8$, (d) $c=0.9$.}
    \label{bb_c}
\end{figure}

We first examine temperature dependence when the gluon condensate
is absent. Figure \ref{sp_t} shows charmonium and bottomonium
spectra at several temperatures for $c=0$. At $T=195$ MeV,
charmonium displays a sharp peak corresponding to the $J/\psi$
ground state, indicating that the state remains relatively stable
and largely intact against thermal dissociation at this
temperature. The bottomonium spectrum features a dominant
$\Upsilon(1S)$ peak accompanied by visible secondary peaks from
excited states such as $\Upsilon(2S)$ and $\Upsilon(3S)$. The
presence of these higher resonances reflects the larger mass of
bottomonium, which makes it more resistant to thermal melting at
moderate temperatures. For both systems, a clear trend emerges:
increasing temperature reduces peak height and broadens
resonances. This is the hallmark of medium-induced dissociation.
Higher temperatures amplify thermal fluctuations in the plasma,
raising the probability that heavy quarkonium bound states break
apart and lowering the spectral weight associated with
quasiparticle resonances.

Next we investigate how the gluon condensate modifies charmonium
spectra at $T=280\,\mathrm{MeV}$ (Fig.~\ref{cc_c}). We select
representative values $c = 0.2,0.5,0.8,0.9$ to track the effect of
increasing condensate strength~\cite{YQZ2020,YK2009}. Even well
above $T_c$, the 1S resonance remains clearly distinguishable,
alongside weaker peaks from excited states. Excited-state
resonances are substantially broader and suppressed relative to
the ground state, consistent with their larger spatial extent and
weaker binding, which renders them more susceptible to medium
effects. A noteworthy feature visible across all panels is that
the central energy position of the 1S peak does not shift as $c$
grows. This separation originates from the two independent scalar
sectors in our construction: vacuum meson masses are fixed by the
soft-wall scalar $\chi(z)$ and fitted quark parameters $k_c,k_b$,
while the gluon condensate parameter $c$ only deforms the finite
temperature black hole geometry controlled by $\phi(z)$. Changes
in $c$ affect resonance widths and spectral weights without
altering the real part of the quasiparticle energy that sets the
peak location. The most important trend is the progressive
narrowing and heightening of the 1S peak as $c$ increases.
Physically, this corresponds to suppressed dissociation: a
stronger gluon condensate enhances nonperturbative gluonic
interactions between the heavy quark and antiquark, reinforcing
the binding of the ground state. Thermal fluctuations are less
able to separate the $Q\bar{Q}$ pair, increasing the lifetime
(narrower width) and occupation probability (larger peak height)
of the bound state. Our findings align with studies of the
imaginary interquark potential~\cite{ST2023} and entropic
forces~\cite{ZQZ2020_2}, which both report that the gluon
condensate hinders quarkonium melting. The effect builds up
gradually with increasing $c$, rather than setting in abruptly,
thereby placing nontrivial constraints on models describing the
interactions between the gluon condensate and heavy quarkonia.

To test the generality of this stabilising effect, we turn to
bottomonium spectra at $T=300\,\mathrm{MeV}$ for varying $c$
(Fig.~\ref{bb_c}). At this temperature, the $\Upsilon(1S)$ peak
stays narrow and prominent for all condensate values, again due to
the larger mass and stronger binding of bottomonium compared to
charmonium. As observed for charmonium, the central position of
the $\Upsilon(1S)$ resonance does not shift with $c$, since vacuum
mass scales are decoupled from the dilaton backreaction parameter.
The clearest variations appear in the excited-state peaks, such as
the 2S resonance. As $c$ rises, broadening weakens and peak height
increases; higher excited states behave qualitatively similarly.
These results confirm that the gluon condensate suppresses thermal
dissociation not only for charmonium but also for bottomonium,
independently of the heavy quark mass. We also observe a
mass-dependent difference between the two systems. For the same
increment in $c$, the narrowing and amplification of the
bottomonium 2S peak is more pronounced than for charmonium. This
can be understood by noting that at the temperatures considered,
the binding energy of bottomonium excited states is larger
relative to the thermal energy compared with charmonium excited
states, making them more sensitive to changes in the effective
interaction induced by the gluon condensate. This comparison
represents a new observation that has not been emphasized in
previous separate analyses of charmonium and bottomonium.

\section{Summary and Discussion}
\label{sec:summary} In this work we have studied the impact of the
gluon condensate on heavy quarkonium spectral functions and
thermal dissociation within a holographic QCD framework. The gluon
condensate is a fundamental nonperturbative QCD quantity, related
to the vacuum expectation value of $G_{\mu\nu}^2$, and encodes key
features including spontaneous chiral symmetry breaking and
nontrivial vacuum topology. Heavy quarkonia, formed as bound
$Q\bar{Q}$ systems, act as valuable QGP probes, and their spectral
functions and melting behavior offer direct access to medium
properties. Through numerical scans across temperature and gluon
condensate strength, we find systematic modifications of resonance
profiles. Larger gluon condensates sharpen the ground-state (1S)
spectral peak, signaling increased effective binding and reduced
decay width. Excited-state peaks also become better defined with
reduced broadening, consistent with suppressed transition rates
towards dissociation.

Physically, these spectral signatures demonstrate that the gluon
condensate strengthens nonperturbative gluon-mediated interactions
between heavy quarks, which impedes thermal dissociation inside
hot QGP and stabilizes quarkonium bound states. This picture
receives independent support from two distinct melting mechanisms
discussed in the literature: analysis of imaginary interquark
potentials~\cite{ST2023} and calculations of entropic
forces~\cite{ZQZ2020_2}. Both approaches arrive at the conclusion
that an enhanced gluon condensate improves quarkonium stability in
the plasma.

Our results are also consistent with our earlier CE analysis in
geometries with a gluon condensate~\cite{wang2025}. The sharper
resonances, reduced broadening, and suppressed dissociation seen
in the spectral calculations provide a microscopic realization of
the increased system stability inferred from CE. This mutual
consistency bridges macroscopic entropy-based stability arguments
and microscopic spectral observables, demonstrating that
gluon-condensate-induced stabilization of heavy quarkonia in the
QGP is a robust nonperturbative effect rather than an artifact of
the chosen formalism. This cross-check also reinforces the utility
of holographic QCD for describing quarkonium dynamics and aids
interpretation of heavy-ion experimental measurements.

Looking forward, we plan to compare our spectral functions with
lattice QCD and potential model results to situate our predictions
within the broader landscape of quarkonium theory. We also aim to
confront the model with upcoming experimental data from LHC Run 4
and RHIC Beam Energy Scan II. Such comparisons will help further
constrain the magnitude and phenomenological consequences of the
gluon condensate in hot QCD matter.

\section{Acknowledgments}
This work is supported by the National Natural Science Foundation
of China (NSFC) under grant No.12375140 and the Fundamental
Research Funds for National Universities, China University of
Geosciences No.2025XLB102.

\end{document}